\UseRawInputEncoding
\documentclass[12pt,letter]{article}
\pdfoutput=1
\usepackage{graphicx, epsfig, color,cite,inputenc}
\usepackage{amsmath}
\usepackage{amssymb}
\usepackage{float}
\usepackage{caption,subcaption,graphicx}
\usepackage{hyperref}

\def\bi{\begin{itemize}}
\def\ei{\end{itemize}}

\def\tu{\tilde u}

\def\tst{\tilde t}

\def\tg{\tilde g}

\def\tq{\tilde q}

\def\agt{\gtrsim}
\def\be{\begin{equation}}  
\def\ee{\end{equation}}  
\def\bea{\begin{eqnarray}}  
\def\eea{\end{eqnarray}}  

\def\SUSY{\text{SUSY}}  
\def\EWSB{\text {EWSB}}  
\def\PU{\text {PU}}  
\def\OU{\text {OU}}  
\def\DSB{\text {DSB}}  
\def\SSB{\text {SSB}}  
\def\EW{\text {EW}}  
\def\cc{\text {cc}}  
\def\GC{\text {GC}}

\begin{document}
\begin{titlepage}
\begin{flushright}
OU-HEP-210322
\end{flushright}

\vspace{0.5cm}
\begin{center}
{\Large \bf Sparticle and Higgs boson masses from the landscape:\\
dynamical versus spontaneous supersymmetry breaking
}\\ 
\vspace{1.2cm} \renewcommand{\thefootnote}{\fnsymbol{footnote}}
{\large Howard Baer$^1$\footnote[1]{Email: baer@ou.edu },
Vernon Barger$^2$\footnote[2]{Email: barger@pheno.wisc.edu},
Shadman Salam$^1$\footnote[3]{Email: shadman.salam@ou.edu} 
and
Hasan Serce$^1$\footnote[4]{Email: hasbarser@gmail.com}
}\\ 
\vspace{1.2cm} \renewcommand{\thefootnote}{\arabic{footnote}}
{\it 
$^1$Homer L. Dodge Department of Physics and Astronomy,
University of Oklahoma, Norman, OK 73019, USA \\[3pt]
}
{\it 
$^2$Department of Physics,
University of Wisconsin, Madison, WI 53706 USA \\[3pt]
}

\end{center}

\vspace{0.5cm}
\begin{abstract}
\noindent
Perturbative supersymmetry breaking on the landscape of string vacua is
expected to favor large soft terms as a power-law or log distribution, 
but tempered by an anthropic veto of
inappropriate vacua or vacua leading to too large a value for 
the derived weak scale- a violation of the atomic principle. 
Indeed, scans of such vacua yield a statistical prediction 
for light Higgs boson mass $m_h\sim 125$ GeV with sparticles 
(save possibly light higgsinos) typically beyond LHC reach.
In contrast, models of dynamical SUSY breaking (DSB)-- with a hidden sector 
gauge coupling $g^2$ scanned uniformly-- lead to gaugino condensation 
and a uniform distribution of soft parameters on a log scale.
Then soft terms are expected to be distributed as $m_{soft}^{-1}$ 
favoring small values. 
A scan of DSB soft terms generally leads to $m_h\ll 125$ GeV
and sparticle masses usually below LHC limits. 
Thus, the DSB landscape scenario seems excluded from LHC search results.
An alternative is that the exponential suppression of the weak scale is
set anthropically on the landscape via the atomic principle.
\end{abstract}
\end{titlepage}

\section{Introduction}
\label{sec:intro}

One of the mysteries of nature is the origin of mass scales. 
At least in QCD, we have an answer: the hadronic mass scale can arise when
the gauge coupling evolves to large values such that the fundamental 
constituents, the quarks, condense to bound states. 
From dimensional transmutation, the proton mass can be found 
even in terms of the Planck mass $m_{Pl}$ via 
$m_{proton}\simeq m_{Pl}\exp (-8\pi^2/g^2)$ which gives the right answer 
for $g^2\sim 1.8$. 

Another mass scale begging for explanation is that
associated with weak interactions: $m_{weak}\simeq m_{W,Z,h}\sim 100$ GeV.
In the Standard Model (SM), the Higgs mass is quadratically divergent so one
expects $m_h$ to blow up to the highest mass scale $\Lambda$ for which the 
SM is the viable low energy effective field theory (EFT).
Supersymmetrization of the SM eliminates the Higgs mass quadratic divergences
so any remaining divergences are merely logarithmic\cite{Witten:1981nf,Kaul:1981wp}: the minimal supersymmetric Standard Model, or MSSM\cite{WSS}, 
can be viable up to the GUT or even Planck scales.
In addition, the weak scale emerges as a derived consequence of the visible 
sector SUSY breaking scale $m_{soft}$. 
So the concern for the magnitude of the weak scale is transferred to a
concern for the origin of the soft breaking scale. 
In gravity mediated SUSY breaking models\footnote{In days of yore, 
gauge mediated SUSY breaking (GMSB) models\cite{Dine:1995ag} were associated with dynamical SUSY breaking in that they allowed much lighter gravitinos. In GMSB models, the trilinear soft term $A_0$ is expected to be tiny, leading to too light a Higgs boson mass unless
soft terms are in the 10-100 TeV regime\cite{Arbey:2011ab,Draper:2011aa,Baer:2012uya}. Such large soft terms then lead to
highly unnatural third generation scalars. For this reason, we focus on DSB in a gravity-mediation context\cite{Bose:2012gq}.}, it is popular to impose
spontaneous SUSY breaking (SSB) at tree level in the hidden sector, for instance
via the SUSY breaking Polonyi superpotential\cite{Polonyi:1977pj}: 
$W=m_{hidden}^2 (\hat{h}+\beta )$ where $\hat{h}$ is the lone hidden sector field.
For $\beta =(2-\sqrt{3})m_P$ (with $m_P$ the reduced Planck mass 
$m_P\equiv m_{Pl}/\sqrt{8\pi}$ and $m_{hidden}\sim 10^{11}$ GeV) then one
determines $m_{soft}\sim m_{3/2}\sim m_{weak}$. 
Thus, the exponentially-suppressed hidden sector mass scale must be put 
in by hand, so SSB can apparently only accommodate, 
but not explain, the magnitude of the weak scale.\footnote{A related 
problem is how the SUSY conserving $\mu$ parameter is {\it also}
generated at or around the weak scale. A recent explanation augments the
MSSM by a Peccei-Quinn (PQ) sector plus a $\mathbb{Z}_{24}^R$ 
discrete $R$-symmetry\cite{Lee:2011dya} 
which generates a gravity-safe accidental 
approximate $U(1)_{PQ}$ which solves the strong $CP$ and SUSY $\mu$ problems, 
and leads to an axion decay constant $f_a\sim m_{hidden}$ 
whilst $\mu\sim m_{weak}$\cite{Baer:2018avn}.
A recent review of 20 solutions to the SUSY $\mu$ problem is given
in Ref. \cite{Bae:2019dgg}.}

A more attractive mechanism follows the wisdom of QCD and seeks to
generate the SUSY breaking scale from dimensional transmutation, which
automatically yields an exponential suppression. 
This is especially attractive in string models where the Planck scale is 
the only mass scale available. Then one could arrange for dynamical
SUSY breaking (DSB)\cite{Dimopoulos:1981au,Witten:1981nf,Dine:1981za} 
(for reviews, see \cite{Poppitz:1998vd,Shadmi:1999jy,Dine:2010cv}) 
wherein SUSY breaking arises non-perturbatively.\footnote{The DSB scenario
has been made more plausible in recent years with the advent of {\it metastable}
DSB\cite{Intriligator:2006dd,Dine:2010cv}.}
Some possibilities include hidden sector gaugino condensation\cite{Ferrara:1982qs}, where
a hidden sector gauge group such as $SU(N)$ becomes confining at the scale
$\Lambda_{\GC}$ and a gaugino condensate occurs with 
$\langle\lambda\lambda\rangle\sim \Lambda_{\GC}^3$ leading to SUSY breaking 
with soft terms $m_{soft}\sim \Lambda_{\GC}^3/m_P^2$. 
The associated hidden mass scale\cite{Affleck:1984mf} is given by
\be
m_{hidden}^2\sim m_P^2\exp (-8\pi^2/g_{hidden}^2) 
\ee 
where then $m_{hidden}^2\sim \Lambda_{\GC}^3/m_P$.
Another possibility is non-perturbative SUSY breaking via instanton effects which 
similarly leads to an exponential suppression of mass scales\cite{Affleck:1983rr}.
Of course, now the mass scale selection problem has been transferred to
the selection of an appropriate value of $g_{hidden}^2$.

A solution to the origin of mass scales also arises within 
the string landscape picture\cite{Bousso:2000xa,Susskind:2003kw}. 
This picture makes use of the vast array of
string vacua found in IIB flux compactifications\cite{Douglas:2006es}. 
Some common estimates from vacuum counting\cite{Ashok:2003gk} are 
$N_{vac}\sim 10^{500}-10^{272,000}$\cite{Denef:2004ze,Taylor:2015xtz}. 
The landscape then provides a setting for Weinberg's anthropic solution to the 
cosmological constant problem\cite{Weinberg:1987dv}: 
the value of $\Lambda_{\cc}$ is expected to be as large as possible such that the expansion rate of the early 
universe allows for galaxy condensation, and hence the 
{\it structure formation} that seems essential for the emergence of life. 

Can similar reasoning be applied to the origin of the weak scale, 
or better yet, the origin of the SUSY breaking scale? 
This issue has been explored initially in Ref's \cite{Susskind:2004uv}, 
\cite{Douglas:2004qg} and \cite{ArkaniHamed:2005yv}.
Here, one assumes a fertile patch of the landscape of vacua where the MSSM is 
the visible sector low energy EFT.
The differential distribution of vacua is expected to be of the form
\be
dN_{vac}[m_{hidden}^2,m_{weak},\Lambda_{\cc}] = f_{\SUSY}\cdot f_{\EWSB}\cdot f_{\cc}\cdot dm_{hidden}^2
\ee
where $f_{\SUSY}(m_{hidden}^2)$ contains the distribution of SUSY breaking mass
scales expected on the fertile patch and $f_{EWSB}$ contains the anthropic
weak scale selection criteria. Denef and Douglas have argued that the
cosmological constant selection acts independently and hence does not affect
landscape selection of the SUSY breaking scale\cite{Denef:2004ze}.

For SSB, then SUSY breaking $F_i$- and $D_\alpha$-terms are expected to be 
uniformly distributed across the landscape, the first as complex numbers
and the latter as real numbers\cite{Douglas:2004qg}. 
This would lead, in the case of spontaneous SUSY breaking, 
to a power law distribution of soft terms
\be
f_{\SUSY}^{\SSB}\sim m_{soft}^n
\ee
where $n={2n_F+n_D-1}$ and $n_F$ are the number of hidden sector SUSY 
breaking $F$-fields and $n_D$ is the number of hidden sector $D$-breaking 
fields contributing to the overall SUSY breaking scale. 
Such a distribution would tend to favor
SUSY breaking at the highest possible mass scales for $n\ge 1$.
Also, Broeckel {\it et al.}\cite{Broeckel:2020fdz} analyzed the distributions 
of SUSY breaking scales from vacua for KKLT\cite{Kachru:2003aw} 
and LVS\cite{Balasubramanian:2005zx} flux compactifications and
found for the KKLT model that $f_{\SUSY}\sim m_{soft}^2$ while the LVS model
gives $f_{\SUSY}\sim \log (m_{soft})$\cite{Baer:2020dri}.

For the anthropic selection, an initial guess was to take 
$f_{\EWSB}=(m_{weak}/m_{soft})^2$ corresponding to a simple fine-tuning factor
which invokes a penalty for soft terms which stray too far beyond the measured 
value of the weak scale. As emphasized in Ref. \cite{Baer:2016lpj} and \cite{Baer:2017uvn}, 
this breaks down in a number of circumstances: 1. soft terms leading
to charge-or-color-breaking (CCB) vacua must be vetoed, not just penalized, 
2. soft terms for which EW symmetry doesn't even break also ought 
to be vetoed (we label these as noEWSB vacua),
3. for some soft terms, the larger they get, then the {\it smaller} 
becomes the derived value of the weak scale. 
To illustrate this latter point, 
we write the {\it pocket universe} (PU)\cite{Guth:1999rh} 
value of the weak scale in terms of 
the pocket-universe $Z$-boson mass $m_Z^{\PU}$ and use the
MSSM Higgs potential minimization conditions to find:
\be
(m_Z^{\PU})^2/2=\frac{m_{H_d}^2+\Sigma_d^d -(m_{H_u}^2+\Sigma_u^u)\tan^2\beta}
{\tan^2\beta -1}-\mu^2\simeq -m_{H_u}^2-\Sigma_u^u-\mu^2
\ee
where $m_{H_{u,d}}^2$ are Higgs soft breaking masses, $\mu$ is the superpotential
Higgsino mass arising from whatever solution to the SUSY $\mu$ problem
is invoked, and $\tan\beta\equiv v_u/v_d$ is the ratio of Higgs field vevs.
The $\Sigma_u^u$ and $\Sigma_d^d$ contain over 40 1-loop radiative
corrections, listed in the Appendix of Ref. \cite{rns}.
The soft term $m_{H_u}^2$ must be driven to negative values at the weak 
scale in order to break EW symmetry. If its high scale value is small, then
it is typically driven deep negative so that compensatory fine-tuning is needed
in the $\mu$ term. If $m_{H_u}^2$ is too big, then it doesn't even run 
negative and EW symmetry is unbroken. The landscape draw to large soft terms
pulls $m_{H_u}^2$ big enough so EW symmetry barely breaks, corresponding to
a natural value of $m_{H_u}^2$ at the weak scale. 
(this can be considered as a landscape selection mechanism for
tuning the high scale value of $m_{H_u}^2$ to such large values 
that its weak scale value becomes natural.)
Also, for large negative
values of trilinear soft term $A_t$, then large cancellations occur
in $\Sigma_u^u (\tst_{1,2})$ leading to more natural $\Sigma_u^u$ values
and a large $m_h\sim 125$ GeV due to large stop mixing in its 
radiative corrections. Also, large values of first/second generation
soft scalar masses $m_0(1,2)$ cause stop mass soft term running to small 
values, thus also making the spectra more natural\cite{Baer:2019cae}.

The correct anthropic condition we believe was set down by Agrawal, 
Barr, Donoghue and Seckel (ABDS) in Ref. \cite{Agrawal:1997gf}. 
In that work, they show that for variable values of the weak scale, 
then nuclear physics is disrupted
if the pocket-universe value of the weak scale $m_{weak}^{\PU}$ deviates
from our measured value $m_{weak}^{\OU}$ by a factor $2-5$. 
For values of $m_{weak}^{\PU}$ outside this range, then nuclei and hence atoms as
we know them wouldn't form. In order to be in accord with this 
{\it atomic principle}, then to be specific, we require 
$m_{weak}^{\PU}<4 m_{weak}^{\OU}$. In the absence of fine-tuning of $\mu$, 
this requirement is then the  same as requiring 
the electroweak fine-tuning measure\cite{ltr,rns} $\Delta_{\EW}<30$.
Thus, we require 
\be f_{\EWSB}=\Theta (30-\Delta_{\EW})
\label{eq:fewsb}
\ee
as the anthropic condition while also vetoing CCB and noEWSB vacua.

\begin{figure}[tbh]
\begin{center}
\includegraphics[height=0.4\textheight]{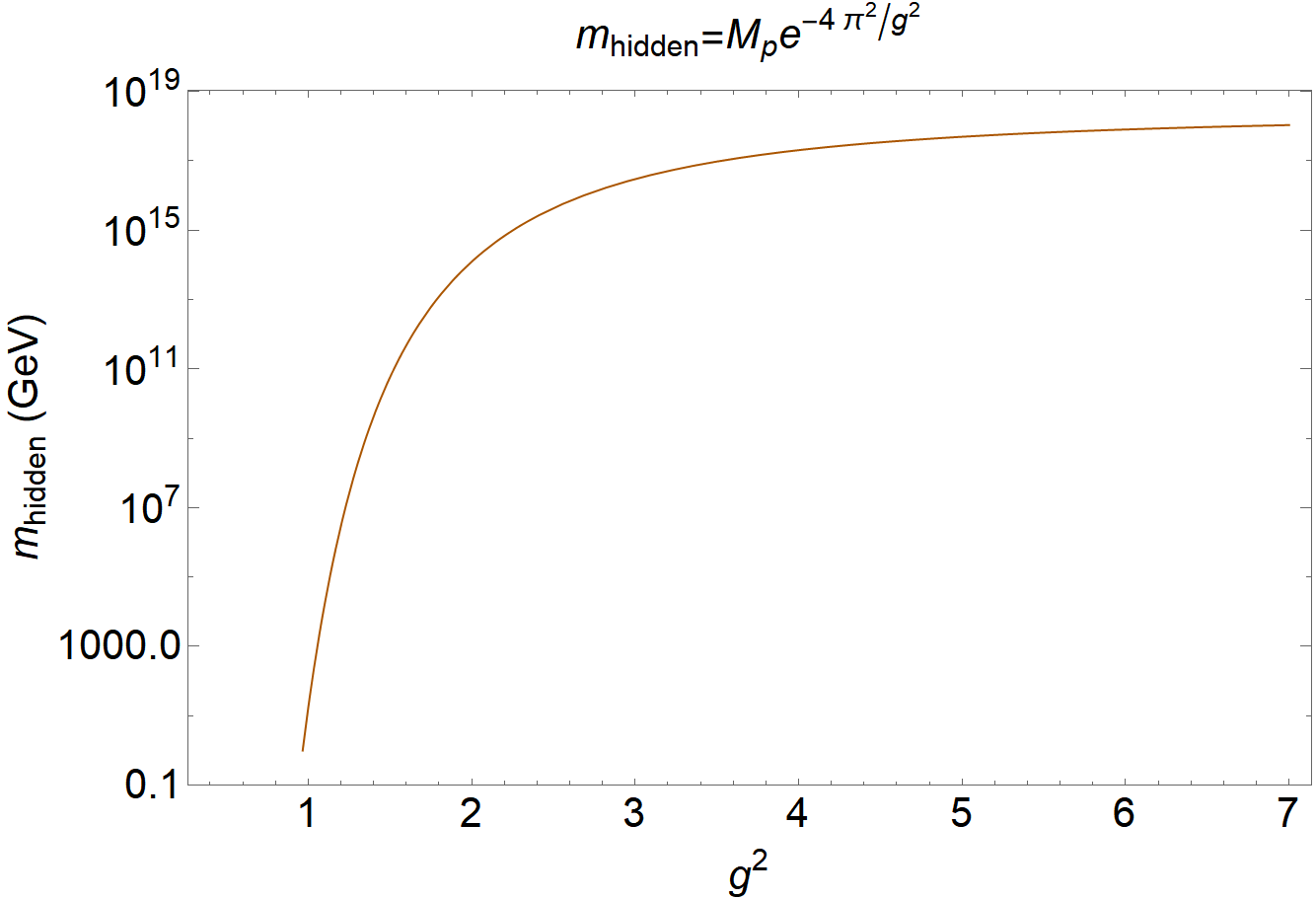}
\caption{Expected SUSY breaking scale $m_{hidden}$ vs. 
hidden sector coupling $g^2$ from dynamical SUSY breaking.
\label{fig:mvsgs}}
\end{center}
\end{figure}

For the case of {\it dynamical} SUSY breaking, the SUSY breaking scale
is expected to be of the form $m_{hidden}^2\sim m_P^2\exp (-8\pi^2/g_{hidden}^2)$ 
where in the case of gaugino condensation, $g_{hidden}$ is the coupling constant of the 
confining hidden sector gauge group.
It is emphasized by Dine {\it et al.}\cite{Banks:2003es,Dine:2004is,Dine:2015xga} and by Denef and Douglas\cite{Denef:2004cf} that the coupling
$g_{hidden}^2$ is expected to scan uniformly on the landscape. 
According to Fig.~\ref{fig:mvsgs}, for $g_{hidden}^2$ values in 
the confining regime $\sim 1-2$, we expect a uniform distribution
of soft breaking terms on a log scale: {\it i.e.} each possible decade
of values for $m_{soft}$ is as likely as any other decade.
Thus, with $m_{soft}\sim m_{hidden}^2/m_P\sim \Lambda_{\GC}^3/m_P^2$, 
we would expect 
\be
f_{\SUSY}^{\DSB}\sim 1/m_{soft}
\ee 
which provides a uniform distribution of $m_{soft}$ across the decades
of possible values\footnote{Dine\cite{Banks:2003es,Dine:2004is,Dine:2005iw,Dine:2015xga} actually finds 
$f_{\SUSY}\sim 1/[m_{soft}\log (m_{soft})]$ which is also highly 
uniform across the decades. We have checked that Dine's distribution 
gives even softer mass distributions than the $1/m_{soft}$ 
which we use.}. 
Such a distribution of course favors 
the {\it lower} range of soft term values.

\section{Results}
\label{sec:results}

Next, we will present the results of calculations of the string landscape
probability distributions for Higgs and sparticle masses under the 
assumption of $f_{\SUSY}^{\DSB}=1/ m_{soft}$ along with Eq. \ref{eq:fewsb} for 
$f_{\text{EWSB}}$. 
Our results will be presented within the gravity-mediated three extra parameter
non-universal Higgs model NUHM3 with parameter space given 
by\cite{nuhm2,nuhm22,nuhm23,nuhm24,nuhm25,nuhm26}
\be
m_0(1,2),\ m_0(3),\ m_{1/2},\ A_0,\ \tan\beta,\ \mu,\ m_A\ \ \ \ \text{(NUHM3)}.
\ee
We adopt the Isajet\cite{isajet} code for calculation of the Higgs and superparticle
mass spectrum\cite{Baer:1994nc} based on 2-loop RGE running\cite{Martin:1993zk} along with sparticle and Higgs masses
calculated at the RG-improved 1-loop level\cite{Pierce:1996zz}.

To compare our results against similar calculations which were presented 
in Ref. \cite{Baer:2017uvn}--but using $f_{\SUSY}=m_{soft}^n$-- 
we will scan over the same parameter space
\begin{itemize}
\item $m_0(1,2):\ 0.1 - 60$ TeV,
\item $m_0(3):\ 0.1 - 20$ TeV,
\item $m_{1/2}:\ 0.5 - 10$ TeV,
\item $A_0:\ -50 -\ 0$ TeV,
\item $m_A:\ 0.3 - 10$ TeV,
\end{itemize}
using the $f_{\SUSY}^{\DSB}$ distribution for soft terms with $\mu = 150$ GeV 
while $\tan\beta:3-60$ is scanned uniformly.
The goal here was to choose upper limits to our scan parameters
which will lie beyond the upper limits imposed by the anthropic selection from $f_{\text{EWFT}}$.
Lower limits are motivated by current LHC search limits, but also
must stay away from the singularity in the $f_{\SUSY}^{\DSB}$ distribution. 
Our final results will hardly depend on the chosen value of $\mu$ so long as 
$\mu$ is within an factor of a few of $m_{W,Z,h}\sim 100$ GeV.
We expect the different classes of soft terms to scan independently as
discussed in Ref. \cite{Baer:2020vad}. 
We will compare the $f_{\SUSY}^{\DSB}$ results against the $f_{\SUSY}^{\SSB}$ 
results from Ref. \cite{Baer:2017uvn} using an $n=2$ power-law draw.

\begin{figure}[H]
\begin{center}
\includegraphics[height=0.22\textheight]{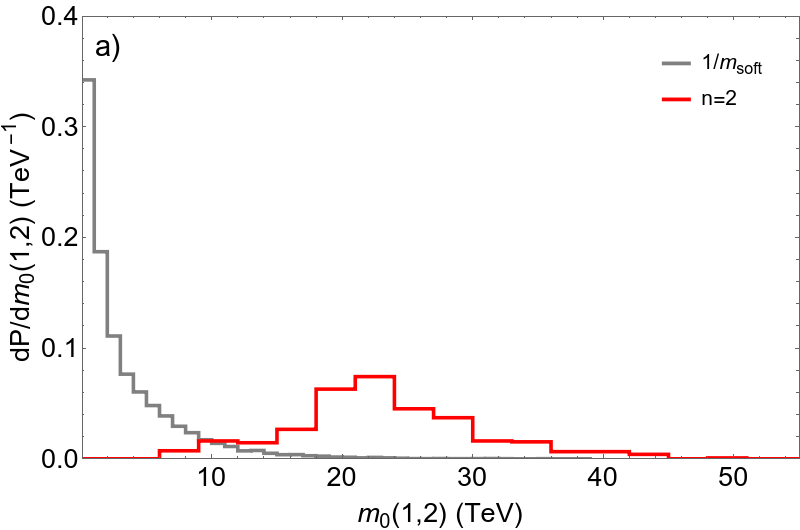}
\includegraphics[height=0.22\textheight]{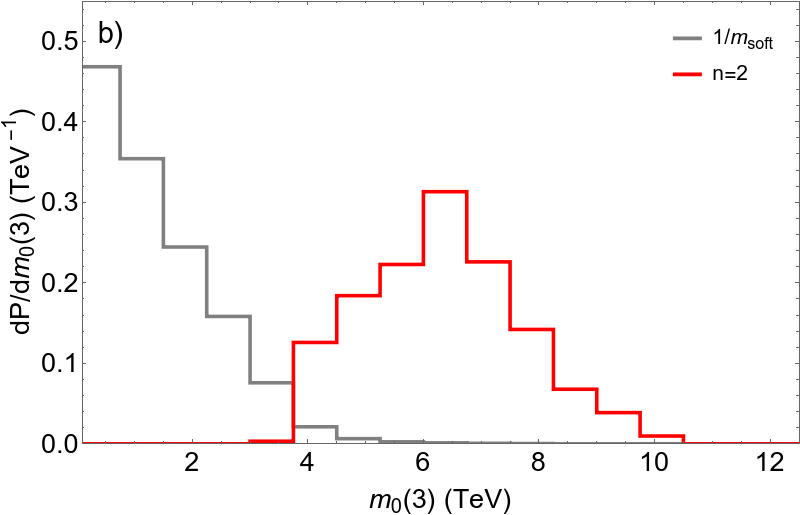}\\
\includegraphics[height=0.22\textheight]{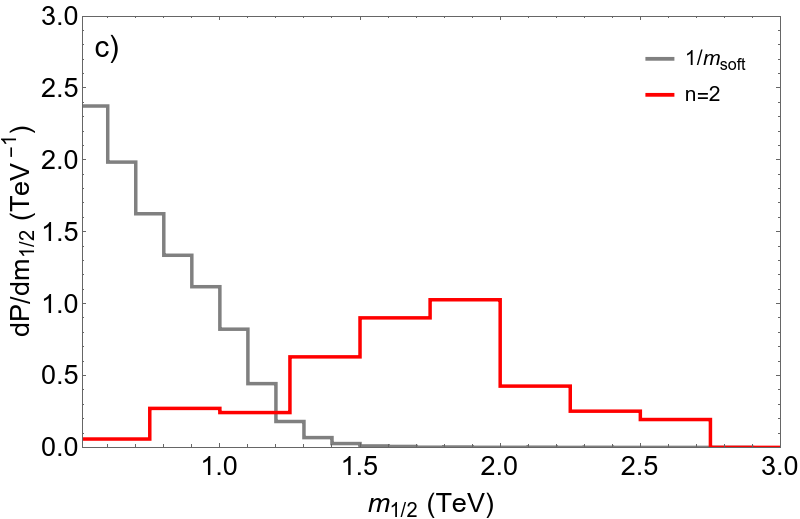}
\includegraphics[height=0.22\textheight]{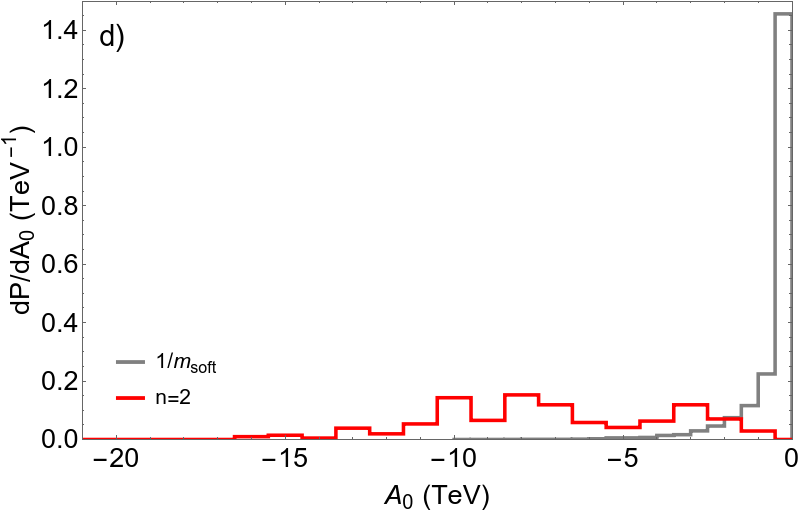}
\caption{Probability distributions for NUHM3 soft terms
{\it a}) $m_0(1,2)$, {\it b}) $m_0(3)$, {\it c}) $m_{1/2}$ and
{\it d}) $A_0$ from a $f_{\SUSY}^{\DSB}=1/m_{soft}$ distribution of soft terms
in the string landscape with $\mu =150$ GeV.
For comparison, we also show probability distributions for 
$f_{\SUSY}^{\SSB}\sim m_{soft}^2$.
\label{fig:m0mhf}}
\end{center}
\end{figure}

In Fig. \ref{fig:m0mhf}, we first show probability distributions for 
various soft SUSY breaking terms for $f_{\SUSY}^{\DSB}$ and also for
$f_{\SUSY}^{\SSB}=m_{soft}^2$. In frame {\it a}), we show the distributions 
versus first/second generation soft breaking scalar masses $m_0(1,2)$.
We see the old SSB $n=2$ result gives a peak distribution at 
$m_0(1,2)\sim 25$ TeV with a tail extending to over 40 TeV. 
This distribution reflects the mixed decoupling/quasi-degeneracy 
landscape solution to the SUSY flavor and CP problems\cite{Baer:2019zfl}.
In contrast, the distribution from $f_{\SUSY}^{\DSB}$ peaks at the lowest 
allowed $m_0(1,2)$ values albeit with a tail extending out beyond
10 TeV. Thus, we would expect relatively light, LHC accessible, 
squarks and sleptons from gravity-mediation with DSB in a hidden sector.
In frame {\it b}), we show the distribution in third generation 
soft mass inputs: $m_0(3)$. Here also the soft terms peak at the lowest values, but this time the tail extends only to $\sim 4$ TeV (lest $\Sigma_u^u (\tst_{1,2})$ becomes too large).
In contrast, the SSB $n=2$ distribution peaks around 7 TeV.
In frame {\it c}), the distribution in unified gaugino soft term 
$m_{1/2}$ is shown. Here again, gaugino masses peak at the lowest allowed scales
for DSB while the $n=2$ distribution peaks just below 2 TeV.
Finally, in frame {\it d}), we show the distribution in trilinear soft term
$-A_0$. Here, the DSB distribution peaks at $-A_0\sim 0$ leading to
little mixing in the stop sector and consequently lower values of 
$m_h$\cite{Carena:2002es,Baer:2011ab}.
In contrast, the $n=2$ distribution has a double peak structure with
peaks at $\sim -4$ and $-7$ TeV with a tail extending to $\sim -15$ TeV:
thus, we expect large stop mixing and higher $m_h$ values from the 
SSB with $n=2$ case.

\begin{figure}[H]
\begin{center}
\includegraphics[height=0.22\textheight]{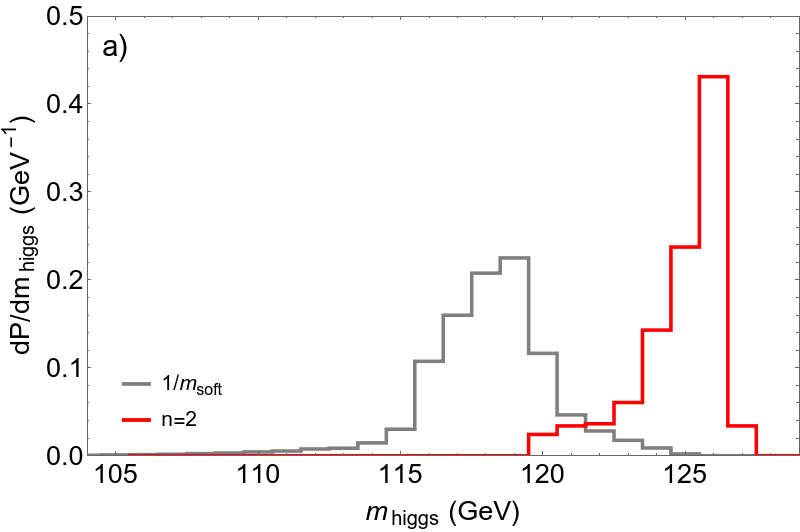}
\includegraphics[height=0.22\textheight]{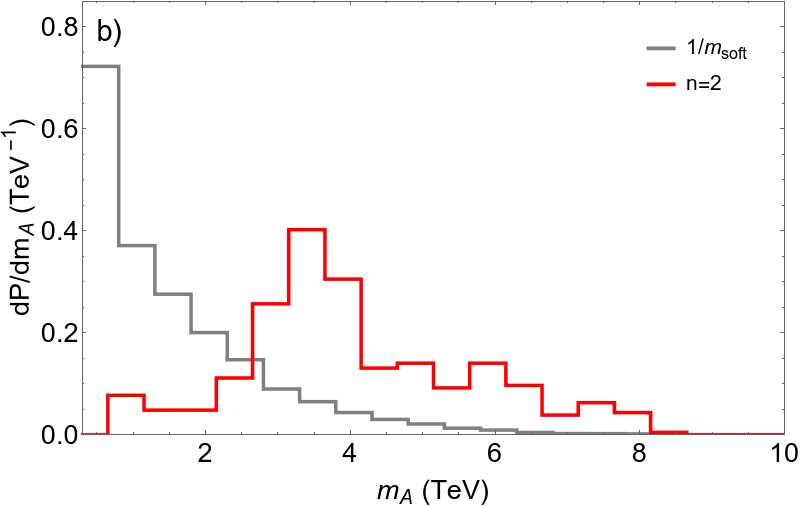}\\
\caption{Probability distributions for light
Higgs scalar mass
{\it a}) $m_h$ and pseudoscalar Higgs mass {\it b}) $m_A$
from a $f_{\SUSY}^{\DSB}\sim 1/m_{soft}$ distribution of soft terms
in the string landscape with $\mu =150$ GeV.
For comparison, we also show probability distributions for 
$f_{\SUSY}^{\SSB}\sim m_{soft}^2$.
\label{fig:higgs}}
\end{center}
\end{figure}

In Fig. \ref{fig:higgs}, we show distributions in light and heavy 
Higgs boson masses. In frame {\it a}), we show the $m_h$ distribution. 
For the DSB case, we see a peak at $m_h\sim 118$ GeV with almost
no probability extending to $\sim 125$ GeV. This is in obvious contrast 
to the data and to the $n=2$ distribution which we see  has a 
sharp peak at $m_h\sim 125-126$ GeV (as a result of large trilinear soft terms).
In frame {\it b}), we see the distribution in pseudoscalar Higgs 
mass $m_A$. In the DSB case, $dP/dm_A$ peaks in the $\sim 300$ GeV range, 
leading to significant mixing in the Higgs sector and consequently 
possibly observable deviations in the Higgs couplings 
(see Ref. \cite{Bae:2015nva}). 
Alternatively, the SSB $n=2$
distribution peaks at $m_A\sim 3.5$ TeV with a tail extending to $\sim 8$ TeV.
In the latter case, we would expect a decoupled Higgs sector with a
very SM-like lightest Higgs scalar $h$ (as indeed the ATLAS/CMS data 
seem to suggest).

\begin{figure}[H]
\begin{center}
\includegraphics[height=0.22\textheight]{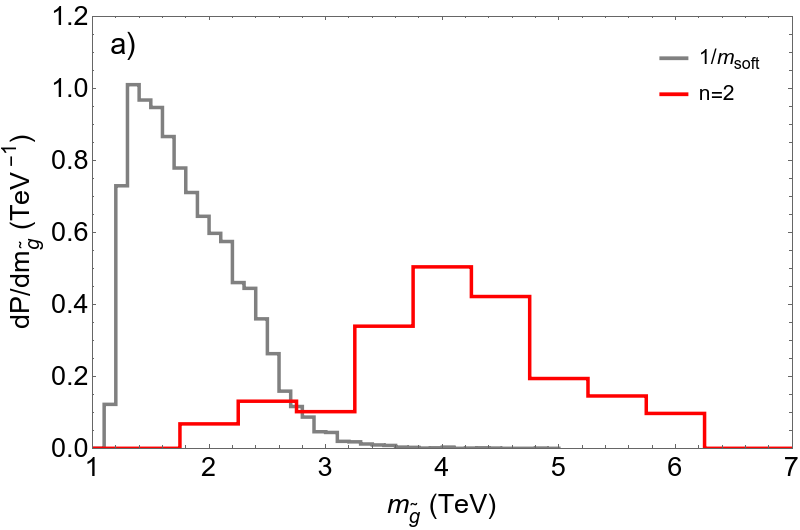}
\includegraphics[height=0.22\textheight]{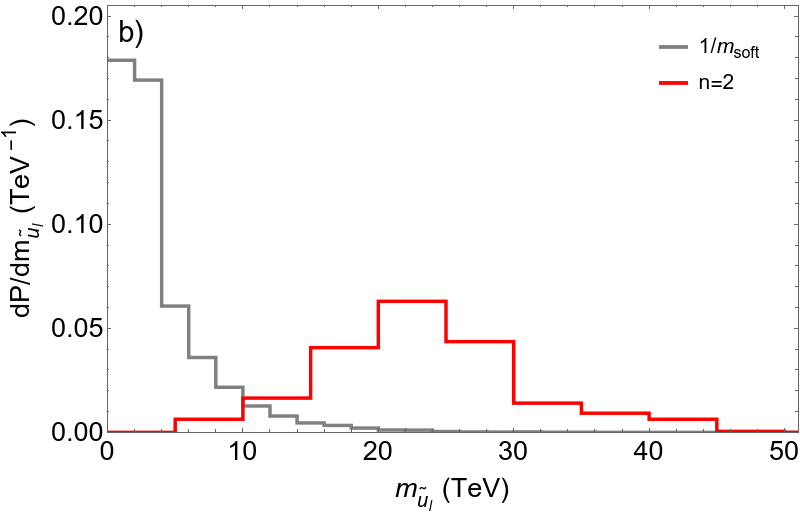}\\
\includegraphics[height=0.22\textheight]{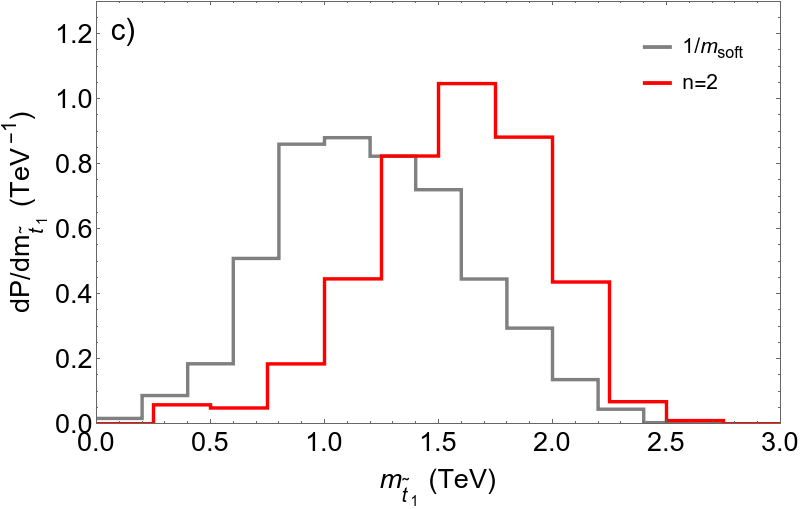}
\includegraphics[height=0.22\textheight]{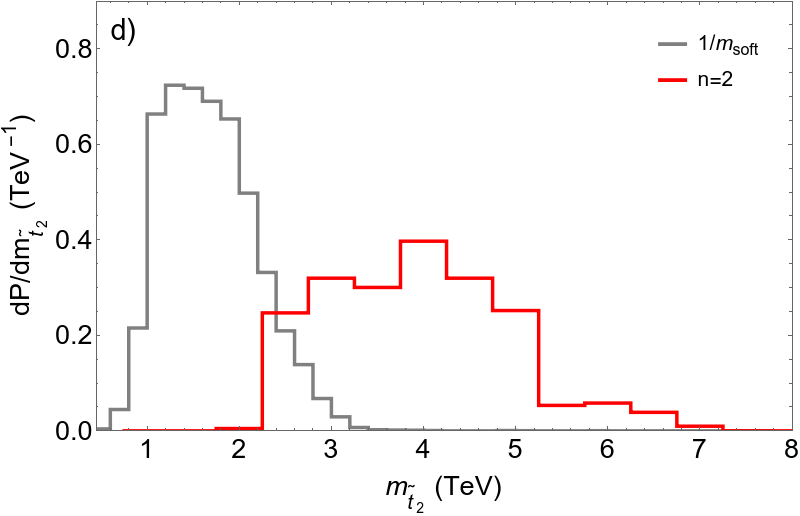}
\caption{Probability distributions for 
{\it a}) $m_{\tg}$, {\it b}) $m_{\tu_L}$, {\it c}) $m_{\tst_1}$ and
{\it d}) $m_{\tst_2}$ from a $f_{\SUSY}^{\DSB}\sim 1/m_{soft}$ distribution of soft terms
in the string landscape with $\mu =150$ GeV.
For comparison, we also show probability distributions for 
$f_{\SUSY}^{\SSB}\sim m_{soft}^2$.
\label{fig:mass}}
\end{center}
\end{figure}

In Fig. \ref{fig:mass}, we show predictions for various sparticle 
masses from the DSB and SSB $n=2$ cases. In frame {\it a}), we show the
distribution in gluino mass $m_{\tg}$. For the DSB case, the distribution peaks
around the $\sim$ TeV range while LHC search limits typically require
$m_{\tg}\agt 2.2$ TeV. In fact, almost all parameter space of DSB is then
excluded. Had we lowered the lower scan cutoff on $m_{1/2}$, the 
distribution would shift lower, making matters worse. The SSB $n=2$
distribution peaks at $m_{\tg}\sim 4-5$ TeV with a tail extending to
$\sim 6$ TeV; hardly any probability is excluded by the LHC $m_{\tg}\agt 2.2$ TeV limit.
In frame {\it b}), we show the distribution in first generation
squark mass $m_{\tu_L}$ (as a typical example of first/second generation matter
scalars). 
The distribution from DSB peaks in the $0-3$ TeV range with a tail 
extending beyond 10 TeV. 
Coupled with the gluino distribution, most probability space would be excluded
by LHC search limits from the $m_{\tg}$ vs. $m_{\tq}$ plane. The SSB
$n=2$ distribution peaks above 20 TeV with a tail extending beyond 40 TeV.
In frame {\it c}), we show the distribution in lighter top squark mass
$m_{\tst_1}$. Here, we see DSB peaks around 1 TeV with a tail to $\sim 2.5$ TeV.
LHC searches require $m_{\tst_1}\agt 1.1$ TeV so that about half of probability space is excluded. For the SSB $n=2$ case, the peak shifts to $m_{\tst_1}\sim 1.6$ TeV so the bulk of p-space is allowed by LHC searches.
Finally, in frame {\it d}), we show the distribution in heavier
stop mass $m_{\tst_2}$. The DSB distribution peaks around $\sim 1.5$ TeV whilst the SSB $n=2$ distribution peaks around $4$ TeV. Thus, substantially heavier $\tst_2$ squarks are expected from SSB as compared to DSB. 

\section{Conclusions}
\label{sec:conclude}

One of the mysteries of particle physics is the origin of mass scales, 
especially in the context of string theory where only the Planck scale 
$m_P$ appears. 
Here, we investigated the origin of the weak scale which is presumed to
arise from the scale of SUSY breaking. 
The general framework of dynamical SUSY breaking presents a beautiful 
example of the exponentially suppressed SUSY breaking scale 
(relative to the Planck scale) arising 
from non-perturbative effects such as gaugino condensation or SUSY breaking 
via instanton effects. The SUSY breaking scale from DSB is expected to 
be uniformly distributed on a log scale within a fertile patch of the string
landscape with the MSSM as the low energy EFT. 
In this case, the probability distribution $f_{\SUSY}^{\DSB}\sim 1/m_{soft}$. 
Such a distribution, coupled with the ABDS anthropic window, 
typically leads to Higgs masses $m_h$ well below
the measured 125 GeV value and many sparticles such as the gluino 
expected to lie below existing LHC search limits. 
Thus, the LHC data seem to falsify this approach. 
That would leave the alternative option of spontaneous
SUSY breaking where instead the soft SUSY breaking distribution is expected
to occur as a power law or log distribution. 
These latter cases lead to landscape probability distributions for $m_h$ 
that peak at $m_h\sim 125$ GeV with sparticles typically well beyond 
current LHC reach, but within reach of hadron colliders with 
$\sqrt{s}\agt 30$ TeV. For perturbative, or spontaneous, SUSY breaking, then 
apparently the magnitude of the SUSY breaking scale is set anthropically
much like the cosmological constant is: those vacua with too large a SUSY
breaking scale lead to either CCB or noEWSB vacua, or vacua with such a large
weak scale that it lies outside the ABDS allowed window, in violation of the 
atomic principle.

{\it Acknowledgements:} 

We thank X. Tata for comments on the manuscript.
This material is based upon work supported by the U.S. Department of Energy, 
Office of Science, Office of High Energy Physics under Award Number DE-SC-0009956 and U.S. Department of Energy (DoE) Grant DE-SC-0017647. 

The computing for this project was performed at the OU Supercomputing Center 
for Education \& Research (OSCER) at the University of Oklahoma (OU).


\bibliography{unilog}
\bibliographystyle{elsarticle-num}

\end{document}